\documentclass[letter,longauth]{aa}

\usepackage{graphicx}
\usepackage{placeins}       
\usepackage{etoolbox}       
\usepackage{txfonts}
\usepackage[
   colorlinks=true,
   citecolor=blue,
   linkcolor=blue,
   urlcolor=blue
]{hyperref}
\defcitealias{gaiacollaboration2023}{Gaia Collaboration (2023)}

\makeatletter
\def\aa@@strip{%
  \ifdim\ht\aa@stripbox>\textheight
    \global\setbox\aa@outputbox=\vsplit\aa@stripbox to \textheight
    \dp\aa@outputbox \z@
    \box\aa@outputbox\clearpage
    \expandafter\aa@@strip
  \else
    \dp\aa@stripbox \z@
    \box\aa@stripbox
  \fi}
\newcommand\aa@patchfail{\errmessage{aa.tex: aa.cls longauth patch failed}}
\patchcmd{\aa@maketitle}{\ifnum\aa@nbauthors>10}{\ifnum\aa@nbauthors>999}{}{\aa@patchfail}
\patchcmd{\aa@maketitle}{\ifnum\aa@nbauthors>10}{\ifnum\aa@nbauthors>999}{}{\aa@patchfail}
\patchcmd{\HyOrg@appendix}{\section*{Authors and affiliations}\aaauthorname}{\section*{Affiliations}}{}{\aa@patchfail}   
\makeatother

\begin{document}

\title{SPIRou near-infrared spectropolarimetry of the ultra-hot Jupiter KELT-9b: tentative detection of the planet magnetic field}

\titlerunning{Tentative magnetic field detection in KELT-9b}
\authorrunning{A. Lavail et al.}

\author{
    A.~Lavail\inst{\ref{instIRAP}}\corrauth{astro@lavail.net}
    \and F.~Debras\inst{\ref{instIRAP}}\email{florian.debras@cnrs.fr}
    \and D.~Roge Royo\inst{\ref{instIRAP}}\email{davirog03@gmail.com}
    \and A.~L\'opez Ariste\inst{\ref{instIRAP}}\email{arturo.lopez-ariste@cnrs.fr}
    \and N.~W.~Borsato\inst{\ref{instIRAP}}\email{nicholas.borsato@utoulouse.fr}
    \and J.-F.~Donati\inst{\ref{instIRAP}}\email{jean-francois.donati@cnrs.fr}
    \and P.~Petit\inst{\ref{instIRAP}}\email{pascal.petit@utoulouse.fr}
    \and J.~D.~Alvarado-G\'omez\inst{\ref{instAIP}}\email{julian.alvarado-gomez@aip.de}
    \and L.~Jouve\inst{\ref{instIRAP}}\email{laurene.jouve@utoulouse.fr}
    \and C.~Moutou\inst{\ref{instIRAP}}\email{claire.moutou@cnrs.fr}
    \and A.~Carmona\inst{\ref{instIRAP}}\email{andres.carmona-gonzalez@univ-tlse3.fr}
    \and B.~Klein\inst{\ref{instOxford}}\email{baptiste.klein@physics.ox.ac.uk}
    \and J.~V. Seidel\inst{\ref{instOCA}}\email{julia.seidel@oca.eu}
    \and L.~Arnold\inst{\ref{instCFHT}}\email{larnold@cfht.hawaii.edu}
    \and E.~Artigau\inst{\ref{instIREX}}\email{etienne.artigau@umontreal.ca}
    \and C.~Baruteau\inst{\ref{instIRAP}}\email{clement.baruteau@utoulouse.fr}
    \and E.~Chabrol\inst{\ref{instLIRA}}\email{estelle.chabrol@obspm.fr}
    \and R.~Cloutier\inst{\ref{instmcmaster}}\email{ryan.cloutier@mcmaster.ca}
    \and N.~J.~Cook\inst{\ref{instIREX}}\email{neil.cook@umontreal.ca}
    \and N.~B.~Cowan\inst{\ref{instMcGill}}\email{nicolas.cowan@mcgill.ca}
    \and X.~Delfosse\inst{\ref{instIPAG}}\email{xavier.delfosse@univ-grenoble-alpes.fr}
    \and N.~Heidari\inst{\ref{instIAP}}\email{heidari@iap.fr}
    \and E.~Martioli\inst{\ref{lna},\ref{instIAP}}\email{martioli@iap.fr}
    \and A.~Masson\inst{\ref{instCAB}}\email{amasson.astro@gmail.com}
    \and J.\,P.~Wardenier\inst{\ref{instBern}}\email{joost.wardenier@unibe.ch}
    \and V.~Yariv\inst{\ref{instIPAG}}\email{vincent.yariv@univ-grenoble-alpes.fr}
}

\institute{
    Univ. de Toulouse, CNRS, IRAP, 14 avenue Belin, 31400 Toulouse, France\label{instIRAP}
    \and Leibniz Institute for Astrophysics Potsdam, 14482 Potsdam, Germany\label{instAIP}
    \and Department of Physics, University of Oxford, Oxford OX1 3RH, United Kingdom\label{instOxford}
    \and Laboratoire Lagrange, Observatoire de la Côte d’Azur, CNRS, Université Côte d’Azur, Nice, France\label{instOCA}
    \and Canada France Hawai`i Telescope Corporation (CFHT), 65-1238 Mamalahoa Hwy, Kamuela, HI 96743, USA\label{instCFHT}
    \and Institut Trottier de recherche sur les exoplanètes, Département de Physique, Université de Montréal, Montréal, Québec, Canada\label{instIREX}
    \and LIRA, Observatoire de Paris, Université PSL, Sorbonne Université, Université Paris Cité, CY Cergy Paris Université, CNRS, 92195 Meudon, France\label{instLIRA}
    \and Department of Physics \& Astronomy, McMaster University, Hamilton, ON, Canada\label{instmcmaster}
    \and Department of Physics and Trottier Space Institute, McGill University, 3600 rue University, H3A 2T8 Montreal QC, Canada\label{instMcGill}
    \and Univ. Grenoble Alpes, CNRS, IPAG, 38000 Grenoble, France\label{instIPAG}
    \and Institut d'Astrophysique de Paris, CNRS, UMR 7095, Sorbonne Université, 98 bis bd Arago, 75014 Paris, France\label{instIAP}
    \and Laborat\'{o}rio Nacional de Astrof\'{i}sica, Rua Estados Unidos 154, 37504-364 Itajub\'{a}, Minas Gerais, Brazil\label{lna}
    \and Centro de Astrobiolog\'ia (CAB), CSIC-INTA, Camino Bajo del Castillo s/n, 28692 Villanueva de la Ca\~nada, Madrid, Spain\label{instCAB}
    \and Weltraumforschung und Planetologie, Physikalisches Institut, University of Bern, Gesellschaftsstrasse 6, 3012 Bern, Switzerland\label{instBern}
}

   \date{Received 17 September 2026 / Accepted dd September 2026}

 
  \abstract
   {Exoplanetary magnetic fields govern atmospheric escape and circulation but have eluded detection in close-in planets. In the ultra-hot Jupiter KELT-9b, atmospheric dynamo is predicted to generate atmospheric fields of 400--4000~G.}
   {We search for the circular polarization in atomic lines that a magnetic field would imprint, through the Zeeman effect, on the transmission spectrum of KELT-9b.}
   {We observed three transits with SPIRou at the CFHT, on 07 September 2025 and 03 and 06 July 2026, and applied least squares deconvolution to 1398 atomic lines to map the transit-averaged Stokes~$V$ and null spectra against orbital semi-amplitude $K_{\rm p}$ and systemic velocity $v_{\rm sys}$.}
   {The 2025 transit shows a Stokes~$V$ signature of $V/I_c = 9.3 \times 10^{-5}$ at $K_{\rm p} = 219 \pm 11$~km\,s$^{-1}$, $v = -34.0 \pm 1.2$~km\,s$^{-1}$, close to the expected planetary velocity. Its false-alarm probability is $1.9 \times 10^{-4}$, formally a marginal detection. Both 2026 transits are of comparable sensitivity but neither exhibits a signal at that location, setting $3\sigma$ limits of $4.9$ and $6.0 \times 10^{-5}$; the 2025 amplitude would have been recovered at $4.9\sigma$ and $4.0\sigma$.}
   {The 2025 signature is robust to the tests we applied to it, but qualifies as a marginal detection at best. Spurious signals of comparable strengths occur elsewhere in our Stokes~$V$ and null maps, and we cannot exclude that our marginal detection is excess noise. It is not reproduced ten months later. }


   \maketitle
   \nolinenumbers   
\section{Introduction}

High-resolution spectroscopy has carried the study of exoplanet atmospheres from the first detection of a single species \citep{charbonneau2002} to detailed compositional and dynamical work \citep[e.g.][]{snellen2025,seidel2025}. Ultra-hot Jupiters, whose molecules are largely dissociated into atoms \citep{hoeijmakers2018}, probe a regime with no Solar System analogue. One fundamental property has so far eluded measurement in close-in planets: magnetic fields, which govern atmospheric escape \citep{owen2017,fossati2021,gronoff2020} and shape atmospheric circulation \citep[e.g.][]{dietrich2022,beltz2024,eliaslopez2025,vigano2025}. Radio campaigns seeking magnetospheric emission analogous to Jupiter's aurorae \citep{zarka2007,griessmeier2007} have yielded only tentative signals \citep[e.g.][]{vedantham2020,zhang2025}, until the recent report of auroral radio emission from the directly imaged giant $\beta$~Pictoris~b, implying $B \gtrsim 1.25$~kG \citep{ortizceballos2026}. Planetary fields have nonetheless been inferred indirectly: from the extended ion tail of HAT-P-11b in ultraviolet transits \citep{benjaffel2022}; from star--planet interactions \citep{pineda2023,revilla2026,zapaterosorio2026}; and, for seven ultra-hot Jupiters, from winds that slow with temperature as magnetic drag predicts, implying internal, dynamo-generated fields of at most a few gauss \citep{seidel2026}. None of these methods measures the Zeeman effect in the planet itself. Fields are however measured in the brown dwarf regime, through radio aurorae \citep{kao2018} and the auroral thermal inversion recently seen with JWST \citep{faherty2024}.

A direct alternative is high-resolution spectropolarimetry, which probes magnetic fields through the Zeeman effect: a field along the line of sight imprints circular polarization (Stokes~$V$) on magnetically sensitive lines. The technique is well established for stellar magnetic fields \citep{donati2009,reiners2012,kochukhov2021}, and exoplanetary atmospheres are predicted to yield detectable signatures \citep{oklopcic2020,strugarek2025}.

Here we apply it to three transit observations of KELT-9b, the hottest known exoplanet. Using SPIRou \citep{donati2020} at the Canada-France-Hawaii Telescope, we combine hundreds of atomic lines by least squares deconvolution (LSD) and recover a candidate marginal circular-polarization signature in one transit out of three.

\section{The KELT-9 system}
\label{sec:system}

KELT-9 (HD~195689) is a bright, rapidly rotating B9.5--A0 main-sequence star \citep{gaudi2017} of $450^{+140}_{-130}$~Myr \citep{borsa2019}. Its planet KELT-9b is among the most extreme known: an equilibrium temperature of $\sim$4050~K, an inflated radius, and a near-polar orbit \citep{gaudi2017}, which together make it highly favourable for transmission spectroscopy. We summarize in Table~\ref{tab:system_parameters} the parameters relevant to this work.

High-resolution spectroscopy has revealed a large inventory of neutral and ionised species in the planet's atmosphere, from \ion{H}{i}, \ion{Na}{i}, and \ion{Mg}{i} to \ion{Fe}{i}, \ion{Fe}{ii}, \ion{Sr}{ii}, OH, and H$^-$ \citep{hoeijmakers2018,hoeijmakers2019,wyttenbach2020,borsato2023,darpa2024,borsato2026}, and evidence for supersonic day-to-night winds of $11.7 \pm 0.6$~km\,s$^{-1}$ on the emerging dayside, exceeding the planetary rotation velocity of $6.4 \pm 0.1$~km\,s$^{-1}$ \citep{zhang2026}. A list of the detections reported for this planet is curated in the ExoAtmospheres database of the Instituto de Astrof\'isica de Canarias\footnote{\url{https://research.iac.es/proyecto/exoatmospheres/view.php?name=KELT-9+b}}. Induction in such an atmosphere is predicted to generate fields of 400--4000~G in the upper layers \citep{dietrich2022}, making KELT-9b, whose host star shows no detected field (Appendix~\ref{app:stellar}), an optimal target for a first search for polarimetric signatures of planetary magnetism.

\section{Observations and data reduction}
\label{sec:observations}

\subsection{SPIRou observations}

We observed three transits of KELT-9b, on 07 September 2025, 03 July 2026, and 06 July 2026, with the SPIRou near-infrared spectropolarimeter \citep{donati2020} at the Canada-France-Hawaii Telescope (CFHT), which covers 965--2450~nm in 49 orders at a resolving power $R \approx 70\,000$ and records intensity and polarized spectra simultaneously. Each polarimetric sequence consisted of four sub-exposures of 111.4~s taken at different polarimeter positions. Demodulation yields an intensity spectrum (Stokes~$I$), a circularly polarized spectrum (Stokes~$V$), and a null spectrum ($N$) built so as to cancel any real polarization, which controls for spurious instrumental signals.

We obtained 33, 32, and 33 sequences on the three nights, respectively, of which 23, 24, and 22 fell within the transit, under comparable conditions: median signal-to-noise ratios of 80.0, 76.8, and 76.8, median seeing $0.55\arcsec$, $0.55\arcsec$, and $0.56\arcsec$, and airmass below 1.4. The observing log is summarised in Table~\ref{tab:obslog} and Fig.~\ref{fig:obslog}.

All observations were taken as part of the PLANETS Large programme (PI: Donati), under run IDs 25BP45 and 26AP45, with the data available at the Canadian Astronomy Data Centre\footnote{\url{https://www.cadc-ccda.hia-iha.nrc-cnrc.gc.ca}} at the end of the proprietary period, in February 2028. The reduced spectra, LSD profiles, analysis code, and line mask needed to reproduce the detection are available now on Zenodo\footnote{\url{https://doi.org/10.5281/zenodo.21806418}}.

\subsection{Data reduction}
\label{sec:reduction}

The raw observations were processed with \texttt{APERO} \citep{cook2022}, the SPIRou data reduction pipeline (v0.7.295), which also corrects telluric absorption and the barycentric Earth radial velocity. Orbital phases and planetary velocities follow the ephemeris of \citet{gaudi2017} and were computed with \texttt{batman} \citep{kreidberg2015}; only the sequences falling inside the transit were retained.

The spectra were shifted to the planetary rest frame, interpolated onto a common grid, averaged over the transit weighted by the squared signal-to-noise ratio of each sequence, and passed to the LSD analysis, once for each trial orbital semi-amplitude $K_{\rm p}$. The three transits were reduced and analysed independently.

\subsection{Least squares deconvolution}
\label{sec:lsd}
\begin{figure*}
   \sidecaption
   \includegraphics[width=0.7\textwidth]{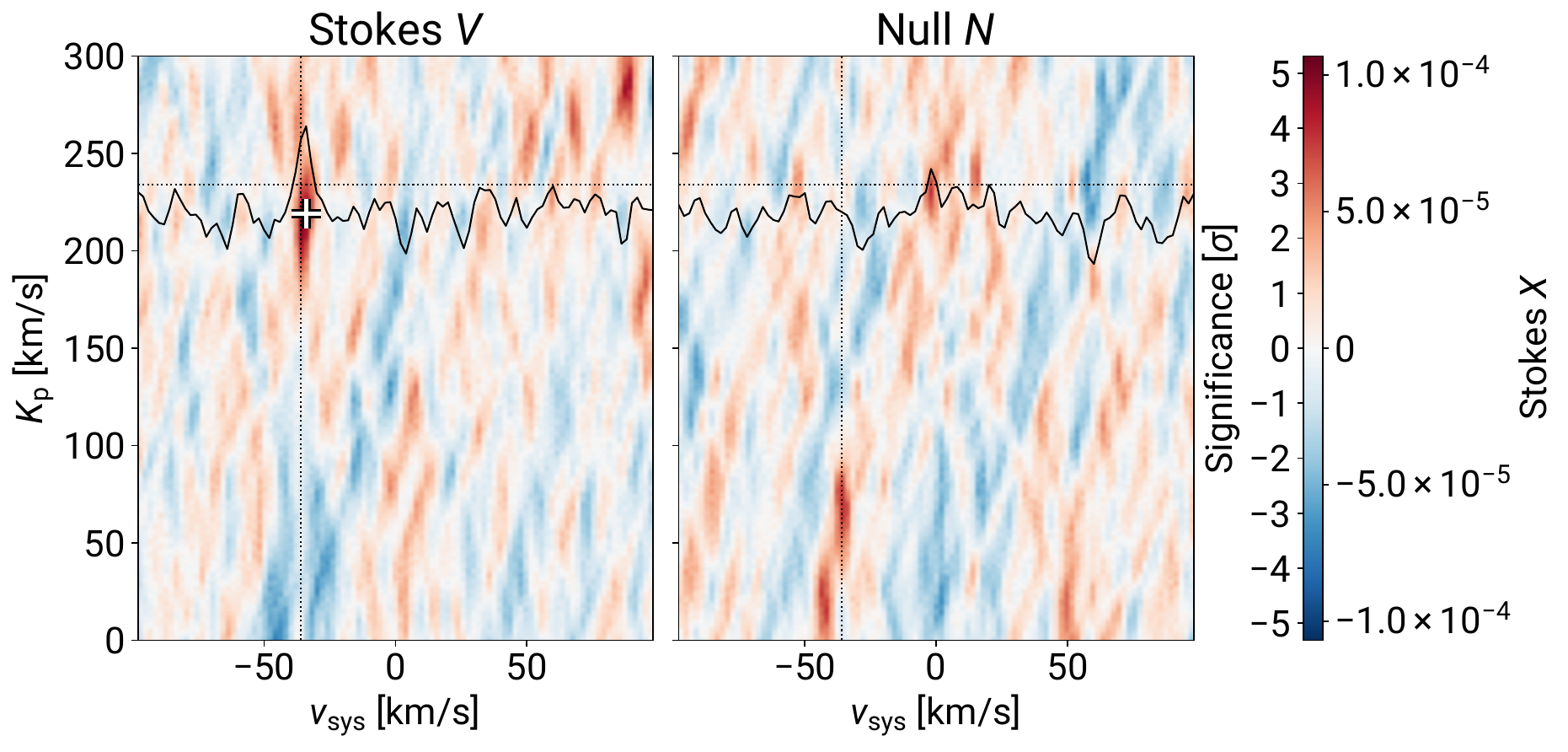}
   \caption{$K_{\rm p}$--$v_{\rm sys}$ maps of the transit-averaged Stokes~$V$ (left) and null~$N$ (right) LSD profiles of KELT-9b, for the transit of 07 September 2025; the three transits are compared in Fig.~\ref{fig:kpmapall}. The colour bars give the significance (left axis) and the corresponding $V/I_c$ or $N/I_c$ (right axis), and the black curves the profiles at the $K_{\rm p}$ of maximum Stokes~$V$. The cross marks the peak quoted in Sect.~\ref{sec:results}; the dotted lines the expected $K_{\rm p} = 234$~km\,s$^{-1}$ \citep{hoeijmakers2019} and velocity $-36$~km\,s$^{-1}$ expected if a single limb contributes (systemic, planetary rotation, and day-to-night wind, Sect.~\ref{sec:disc_velocity}).}
   \label{fig:kpmap}
\end{figure*}
We employed LSD \citep{donati1997,kochukhov2010}, which models the observed spectrum as the convolution of a line mask, holding the wavelength, depth, and Land\'e factor of each line, with a mean profile, to combine hundreds of lines into a single high signal-to-noise profile. The mask was extracted from the Vienna Atomic Line Database \citep[VALD3;][]{ryabchikova2015} for an atmosphere with $T_{\rm eff} = 4000$~K, the conditions near the base of the temperature profile of \citet{fossati2021}, where weak near-infrared lines form in transmission, and $\log g = 3.5$, the nearest node of the extraction grid. We kept atomic lines between 1200 and 2500~nm with a known effective Land\'e factor and a central depth deeper than 10\% of the continuum, excluding H, He, Na, K, and Ca, whose lines can appear in emission. The mask contains 1398 lines from 14 species, dominated by \ion{Fe}{i} with 959 lines, with mean central depth $\langle d \rangle = 0.25$, effective Land\'e factor $\langle g_{\rm eff} \rangle = 1.27$, and wavelength $\langle \lambda \rangle = 1747$~nm.

Each line $i$ was weighted by $w_i = d_i \lambda_i g_{{\rm eff},i} / (\langle d \rangle \langle \lambda \rangle \langle g_{\rm eff} \rangle)$, following \citet{kochukhov2010}. For each trial $K_{\rm p}$ from 0 to 300~km\,s$^{-1}$ in steps of 1~km\,s$^{-1}$ we computed LSD profiles of the transit-averaged Stokes~$V$ and null~$N$ spectra, giving for each transit a map against $K_{\rm p}$ and systemic velocity $v_{\rm sys}$ (from $-400$ to $400$~km\,s$^{-1}$ in steps of 2~km\,s$^{-1}$), and its null counterpart. Significances are quoted in units of the standard deviation of the Stokes~$V$ map over the line-free range $|v| > 200$~km\,s$^{-1}$, and the null maps are scaled by the same quantity so that the two can be compared directly.

\section{Results}
\label{sec:results}

The $K_{\rm p}$--$v_{\rm sys}$ map of the transit of 07 September 2025 is shown in Fig.~\ref{fig:kpmap}. The Stokes~$V$ map peaks at $K_{\rm p} = 219 \pm 11$~km\,s$^{-1}$ and $v = -34.0 \pm 1.2$~km\,s$^{-1}$, with a peak height of $5.3\sigma$ and an amplitude of $V/I_c = 9.3 \times 10^{-5}$.  This position is consistent with the expected orbital semi-amplitude and systemic velocity, once the planetary rotation and the day-to-night wind are accounted for. The strongest value of the corresponding null map reaches $3.9\sigma$, at a different position in the map, and the LSD profiles extracted at the $K_{\rm p}$ of maximum detection are shown in Fig.~\ref{fig:Vprofile}. The false-alarm probability (FAP; Appendix~\ref{app:robustness}) of the signature is $1.9 \times 10^{-4}$ for Stokes~$V$ and $0.71$ for the null, formally constituting a marginal detection. 

The two transits observed in July 2026 were processed and analysed with the same procedure. Neither yields a significant Stokes~$V$ signal; their $K_{\rm p}$--$v_{\rm sys}$ maps are shown in Fig.~\ref{fig:kpmapall}. At the position of the 2025 signature their FAPs are $0.54$ and $0.19$, and both maps are consistent with zero there. The noise at that $K_{\rm p}$, measured over $|v| > 200$~km\,s$^{-1}$, is $1.6 \times 10^{-5}$ on 03 July and $2.0 \times 10^{-5}$ on 06 July, in comparison to $1.8 \times 10^{-5}$ for the 2025 transit; the 2026 observations are thus of comparable sensitivity. They set $3\sigma$ upper limits of $4.9$ and $6.0 \times 10^{-5}$ on any circular polarization at that position: a signal of the amplitude measured in 2025 would have been recovered at $4.9\sigma$ and $4.0\sigma$ (Appendix~\ref{app:robustness}).

The 2026 data therefore do not reproduce the 2025 signature at the amplitude at which it was measured. The 2025 signature does, however, survive the tests we applied to it (Appendix~\ref{app:robustness}): it is recovered at the same location by two independent data reductions and by two different LSD codes. It weakens when the out-of-transit sequences are included, and a split of the mask at its median wavelength recovers it at $4.8\sigma$ from the red half and $1.9\sigma$ from the blue. We therefore regard it as a candidate signature. The 2026 non-detections could be compatible with a signature that varies between epochs, if the 2025 detection can be confirmed with further observations. 

\section{Discussion}
\label{sec:discussion}

The Stokes~$V$ signal detected in 2025, if coming from the planet, should satisfy three constraints. These are the amplitude of the signature ($\sim$$10^{-4}\,I_c$), its position in velocity ($-34.0 \pm 1.2$~km\,s$^{-1}$), and its single-lobed shape.

\subsection{The velocity of the signature}
\label{sec:disc_velocity}

During the transit, the planetary atmosphere is seen against the stellar disc as a thin annulus, and the two limbs contribute with opposite line-of-sight rotation velocities. In a geometry symmetric about the sub-stellar meridian, the rotational contributions of the morning and evening limbs are both present, and the planetary rotation does not displace the centroid of the resulting Stokes~$V$ profile: it broadens it, up to magnetic polarity considerations. The expected position of the signature is then set by the systemic velocity and the day-to-night wind alone, $-17.86 - 11.7 = -29.6$~km\,s$^{-1}$, 4.4~km\,s$^{-1}$ redwards of the observed peak.

However, published systemic velocities of KELT-9 span $-17.2$ to $-20.6$~km\,s$^{-1}$ \citep{asnodkar2022}. A homogeneous analysis of 13 transits gives $-17.99 \pm 0.04$~km\,s$^{-1}$ \citep{borsato2026}, $0.13$~km\,s$^{-1}$ from the value adopted here. The wind is the weaker term: the $11.7$~km\,s$^{-1}$ of \citet{zhang2026} is measured on the dayside in emission, whereas the transit probes the terminator, where \citet{darpa2024} report blueshifts of at most $-11$ and typically $-6$~km\,s$^{-1}$; the latter would put the expectation at $-23.9$~km\,s$^{-1}$, 10~km\,s$^{-1}$ from the observed peak. \citet{borsato2026} also measure a blueshift of $-8.02 \pm 0.17$~km\,s$^{-1}$ for the species detected in transmission, putting the symmetric expectation at $-26.0$~km\,s$^{-1}$, 8~km\,s$^{-1}$ from the peak. Reproducing the observed velocity with a symmetric geometry requires winds of $-16.1$~km\,s$^{-1}$, stronger than any of these. Alternatively, the optical wind measurements and our near-infrared mask could not probe the same depth, where the wind speed may differ.

A possible alternative is to break the east--west symmetry. If only one limb contributes appreciably, the rotational contribution no longer cancels and the centroid is displaced by several km\,s$^{-1}$. The observed velocity then follows without invoking stronger winds: it is consistent at $1.6\sigma$ with $-35.96$~km\,s$^{-1}$, the sum of the systemic velocity, the planetary rotation, and the day-to-night wind. The $\pm1.2$~km\,s$^{-1}$ on the measured velocity is statistical only. The same planet's signature shifts by 5 to 10~km\,s$^{-1}$ between instruments and between species \citep{borsato2024}, more than the 6.4~km\,s$^{-1}$ of planetary rotation that separates the two geometries, so the velocity cannot choose between them. The one-limb case also has a difficulty of its own: it predicts a Stokes~$V$ profile with two lobes of equal strength (Fig.~\ref{fig:fwdlimbs}), whereas the observed signature has a single lobe.

The measured $K_{\rm p}$ leaves the question open as well. At $219 \pm 11$~km\,s$^{-1}$ it lies 1.4--$2.0\sigma$ below the 234--242~km\,s$^{-1}$ obtained from Fe lines in transit and in dayside emission \citep[compiled by][]{asnodkar2022}, $2.5\sigma$ below the $247.2 \pm 1.1$~km\,s$^{-1}$ of \citet{borsato2026}, and $2.7\sigma$ below the circular-orbit value of $254^{+8}_{-7}$~km\,s$^{-1}$, computed from Table~\ref{tab:system_parameters}. A limb contribution that moves from the redshifted morning side to the blueshifted evening side through the transit would lower the apparent $K_{\rm p}$ by tilting the signal trail \citep{wardenier2021,prinoth2022}. The dayside value is low as well, however, so the gap may instead reflect the stellar mass.

\subsection{Scenarios and the implied field strength}
\label{sec:disc_scenarios}

We explored a set of forward toy-models in which the planet is represented by illuminating the limbs of a disc, and setting a purely toroidal magnetic field whose strength at a point on the planetary surface is expressed as $B_{\rm eq}\sin\theta$ ($B_{\rm eq}$ is the magnetic field strength at the equator, $\theta$ is the colatitude of the considered point). We adopted the planetary rotation velocity of \citet{zhang2026} (Appendix~\ref{app:forward}). 

Illuminating both limbs reproduces the amplitude of the signature with $B_{\rm eq} \approx 2$~kG. Illuminating a single limb moves the first moment of $|V|$ to $-34.9$~km\,s$^{-1}$. It predicts, however, a two-lobed profile, where the observed signature is single-lobed. Regarding the velocity of the signals, neither reproduces the observations well (Fig.~\ref{fig:fwdlimbs}). Both are formally inconsistent with the 2025 profile.

A separate difficulty concerns the ratio $V/I$. Extrapolating the Stokes $I$ signature of the planetary lines from the best optical models gives an amplitude of $\sim 10^{-4}$. The intensity profile is not detected in our near-infrared data. Injection--recovery tests on the 2025 transit set an upper limit of $3 \times 10^{-4}$ on $I$ (Appendix~\ref{app:robustness}). Our data therefore require $V/I \gtrsim 0.3$, which is extreme and unobserved in stellar magnetic fields, though the toy model reaches $V/I = 0.5$ once the splitting exceeds the line width (Appendix~\ref{app:forward}).  

Lastly, the single-limb case could be made to produce a single-lobed Stokes~$V$ profile. Single-lobed Stokes~$V$ signatures are actually observed in stars -- e.g. ultra-weak magnetic A stars, asymptotic giant branch stars, or the Sun \citep{folsom2018,lavail2025,lopezariste2001}, a shape attributed to sharp vertical gradients in velocity and field strength across the line-forming region. A magnetic canopy with a velocity or magnetic field gradient present in the atmosphere of KELT-9b is a further possibility to explain our signal.

\subsection{Temporal variability of an induced field}
\label{sec:disc_variability}

The non-detection of the signature in the two July 2026 transits requires either that the signature is not astrophysical, a plausible scenario, or that it varies between epochs. The field expected in KELT-9b's atmosphere is not predicted to be steady.
\citet{dietrich2022} showed that, for planets with $T_{\rm eq} > 1500$~K, the thermal ionisation of metals raises the electrical conductivity to $\sim$1~S\,m$^{-1}$ throughout the atmosphere and makes the induction of atmospheric magnetic fields a runaway process that can only be halted by non-linear feedback: the back-reaction of the Lorentz force on the flow, or the onset of magnetic instabilities. For KELT-9b they conclude that ``the time variability of the induced field remains a key contribution''. The induced fields are predominantly horizontal (i.e. toroidal), as assumed in our forward models, and saturate at up to 4~kG, this value falling to 400~G if the back-reaction of the magnetic field on the shear is taken into account. 

Our rough estimate of $B_{\rm eq} \sim 2$~kG concerns the large scale magnetic field, as small-scale fields suffer cancellation in circular polarization. The total magnetic field is in reality larger. In stars, at best, $\sim$26\% of the total field is recovered by Stokes~$V$ \citep{kochukhov2021}. The total magnetic field would then be stronger than 7.7~kG, above the 4~kG saturation predicted by \citet{dietrich2022}. These induced fields are distinct from the internal, dynamo-generated fields constrained by \citet{seidel2026}. Adopting a large scale magnetic field of 400~G in our forward toy-model gives $V/I_c = 1.8 \times 10^{-5}$, which, given the noise and response of our data (Appendix~\ref{app:robustness}), would take some fifteen to seventeen transits to detect at $3\sigma$. 

The relevant timescale is short compared with our observing baseline: the Alfv\'en timescale estimated for KELT-9b, $t_{\rm A} \approx 150$~d, is an order-of-magnitude estimate rather than a period whose phase can be followed, and our 2025 and 2026 transits are separated by $\sim$300~d. Variability of the induced field between the two epochs could therefore be expected, consistent with the temporal variability of magnetically-driven winds found in the magnetohydrodynamic simulations of \citet{rogers2017}. The two July 2026 transits are three nights apart and form a single epoch. 

The case for variability remains qualitative. The field would have to fall from the kG level to below our 2026 limits, or reorient completely, within a single season, a behaviour that has not been observed in stars, where magnetic variability is common but never that extreme. A change between epochs does not have to be solely magnetic: for example, thicker limb hazes would weaken the planetary lines, and with them any Zeeman signature. Finally, the $5.3\sigma$ peak is the amplitude measured on a single pixel. The noise of the maps is correlated over $21$~km\,s$^{-1}$ in $K_{\rm p}$ and $3.6$~km\,s$^{-1}$ in velocity. The mapped region, $|v| \leq 100$~km\,s$^{-1}$, then holds 760 independent elements (out of $3\times10^4$ pixels) and the trials-corrected significance is $3.9\sigma$. Within the window in which the signature was predicted ($K_{\rm p} = 200$--$260$~km\,s$^{-1}$, $v = -40$ to $-25$~km\,s$^{-1}$) it is $4.8\sigma$. Those figures assume Gaussian noise, which these maps do not exactly have: the strongest excursion away from the signature reaches $4.6\sigma$ in the 2025 map, and $5.1\sigma$ and $4.7\sigma$ in the two 2026 maps, where Gaussian noise would predict a largest excursion of $3.2\sigma$. We therefore cannot exclude that the 2025 peak is an excess-noise peak.

\bibliographystyle{aa} 
\bibliography{kelt9} 

\onecolumn   
\let\aasavedclearpage\clearpage\let\clearpage\relax   
\begin{appendix}
\global\let\clearpage\aasavedclearpage   
\nolinenumbers   

\begin{acknowledgements}
Our study is based on data obtained at the CFHT, operated by the CNRC (Canada), INSU/CNRS (France), and the University of Hawaii. The authors wish to recognise and acknowledge the very significant cultural role and reverence that the summit of Maunakea has always had within the indigenous Hawaiian community. This work has made use of the VALD database, operated at Uppsala University, the Institute of Astronomy RAS in Moscow, and the University of Vienna. F.D acknowledges funding from the French National Research Agency (ANR) project ExoATMO (ANR-25-CE49-6598), and would like to warmly thank the Action Thématique Physique Stellaire, Action Spécifique Numérique and Action Thématique ExoSystèmes of  CNRS/INSU cofunded by CEA and CNES for their financial support. N.W.B. acknowledges funding from the ANR (project ANR-24-CE49-3397 ORVET). B.K. acknowledges funding from the European Research Council under the European Union’s Horizon 2020 research and innovation programme (grant agreement No. 865624, GPRV). E.M. acknowledges research productivity grant number 319217/2025-6 awarded by CNPq, Brazil. A.M. acknowledges fundings by the Spanish Ministerio de Ciencia, Innovaci\'on y Universidades (MICIU/AEI/10.13039/501100011033) and NextGenerationEU/PRTR grants CNS2023-144309 and PID2023-150468NB-I00. J.P.W. acknowledges support from the Swiss National Science Foundation under grant 10002706 (PIs: D. Kitzmann, H.J. Hoeijmakers) and from the Canadian Space Agency under grants 24JWGO3A-03 and 25JWGO4A-14.
\end{acknowledgements}

\section{Properties of the KELT-9 system}
\label{app:system}
\begin{table}[!ht]
\small
\caption{Literature properties of the KELT-9 system used in this work.}
\label{tab:system_parameters}
\centering
\begin{tabular}{llll}
\hline\hline
Property & Symbol & Value & Reference \\
\hline
\multicolumn{4}{l}{\textit{Stellar properties}} \\
\hline
Identifiers & & HD~195689, BD$+$39~4221 & \\
Mean $G$ magnitude & $G$ [mag] & $7.5903 \pm 0.0028$ & \citetalias{gaiacollaboration2023} \\
Distance & $d$ [pc] & $207.2 \pm 1.1$ & \citetalias{gaiacollaboration2023} \\
Effective temperature & $T_{\rm eff}$ [K] & $9495 \pm 104$ & \citet{kama2023} \\
Surface gravity & $\log g$ [dex] & $4.17 \pm 0.17$ & \citet{kama2023} \\
Projected rotation velocity & $v \sin i_\star$ [km\,s$^{-1}$] & $114.9 \pm 3.4$ & \citet{kama2023} \\
Systemic velocity\tablefootmark{a} & $v_{\rm sys}$ [km\,s$^{-1}$] & $-17.860 \pm 0.044$ & \citet{asnodkar2022} \\
Age & $\tau$ [Myr] & $450^{+140}_{-130}$ & \citet{borsa2019} \\
\hline
\multicolumn{4}{l}{\textit{Planetary properties}} \\
\hline
Orbital period & $P_{\rm orb}$ [d] & $1.4811235 \pm 0.0000011$ & \citet{gaudi2017} \\
Transit epoch & $T_0$ [BJD$_{\rm TDB}$] & $2457095.68572$ & \citet{gaudi2017} \\
Semi-major axis & $a$ [au] & $0.03462^{+0.00110}_{-0.00093}$ & \citet{gaudi2017} \\
Scaled semi-major axis & $a/R_\star$ & $3.153 \pm 0.011$ & \citet{gaudi2017} \\
Planet mass & $M_{\rm p}$ [M$_{\rm Jup}$] & $2.88 \pm 0.84$ & \citet{gaudi2017} \\
Planet radius & $R_{\rm p}$ [R$_{\rm Jup}$] & $1.891^{+0.061}_{-0.053}$ & \citet{gaudi2017} \\
Orbital inclination & $i$ [$^\circ$] & $86.79 \pm 0.25$ & \citet{gaudi2017} \\
Projected spin-orbit angle & $\lambda$ [$^\circ$] & $-84.8 \pm 1.4$ & \citet{gaudi2017} \\
Planet radial-velocity semi-amplitude, from \ion{Fe}{ii} & $K_{\rm p}$ [km\,s$^{-1}$] & $234.24 \pm 0.90$ & \citet{hoeijmakers2019} \\
Planetary rotation velocity & $v_{\rm rot,p}$ [km\,s$^{-1}$] & $6.4 \pm 0.1$ & \citet{zhang2026} \\
Day-to-night wind velocity\tablefootmark{b} & $v_{\rm wind}$ [km\,s$^{-1}$] & $11.7 \pm 0.6$ & \citet{zhang2026} \\
Equilibrium temperature\tablefootmark{c} & $T_{\rm eq}$ [K] & $4050 \pm 180$ & \citet{gaudi2017} \\
Upper-atmosphere temperature at $10^{-6}$~bar & $T_{\rm upper}$ [K] & $\sim$8500 & \citet{fossati2021} \\
Thermospheric temperature & $T_{\rm therm}$ [K] & $\sim$13\,200 & \citet{wyttenbach2020} \\
\hline
\end{tabular}
\tablefoot{\tablefoottext{a}{The value of \citet{asnodkar2022} obtained with PEPSI. Their Table~3 compiles six determinations spanning $-17.2$ to $-20.6$~km\,s$^{-1}$, so the quoted uncertainty is statistical only.} \tablefoottext{b}{Measured on the dayside in emission, away from transit, and an upper end: terminator blueshifts are smaller \citep{darpa2024}.} \tablefoottext{c}{For zero albedo and full heat redistribution.}}
\end{table}

Table~\ref{tab:system_parameters} lists the stellar and planetary parameters adopted throughout this work. Published systemic velocities of KELT-9 do not agree. The PEPSI value adopted here, $-17.860 \pm 0.044$~km\,s$^{-1}$, is one of six determinations compiled by \citet{asnodkar2022} that span $-17.2$ to $-20.6$~km\,s$^{-1}$, and \citetalias{gaiacollaboration2023} give $-20.23 \pm 0.49$~km\,s$^{-1}$ from Gaia DR3. \citet{borsato2026} find $-17.99 \pm 0.04$~km\,s$^{-1}$, consistent across six spectrographs and 13 transits and the closest to the value adopted here. We treat the spread of these values as a systematic uncertainty on $v_{\rm sys}$ (Sect.~\ref{sec:disc_velocity}).

\FloatBarrier
\section{Observing log}
\label{app:obslog}

\begin{figure}[!ht]
   \sidecaption
   \includegraphics[width=0.630\textwidth]{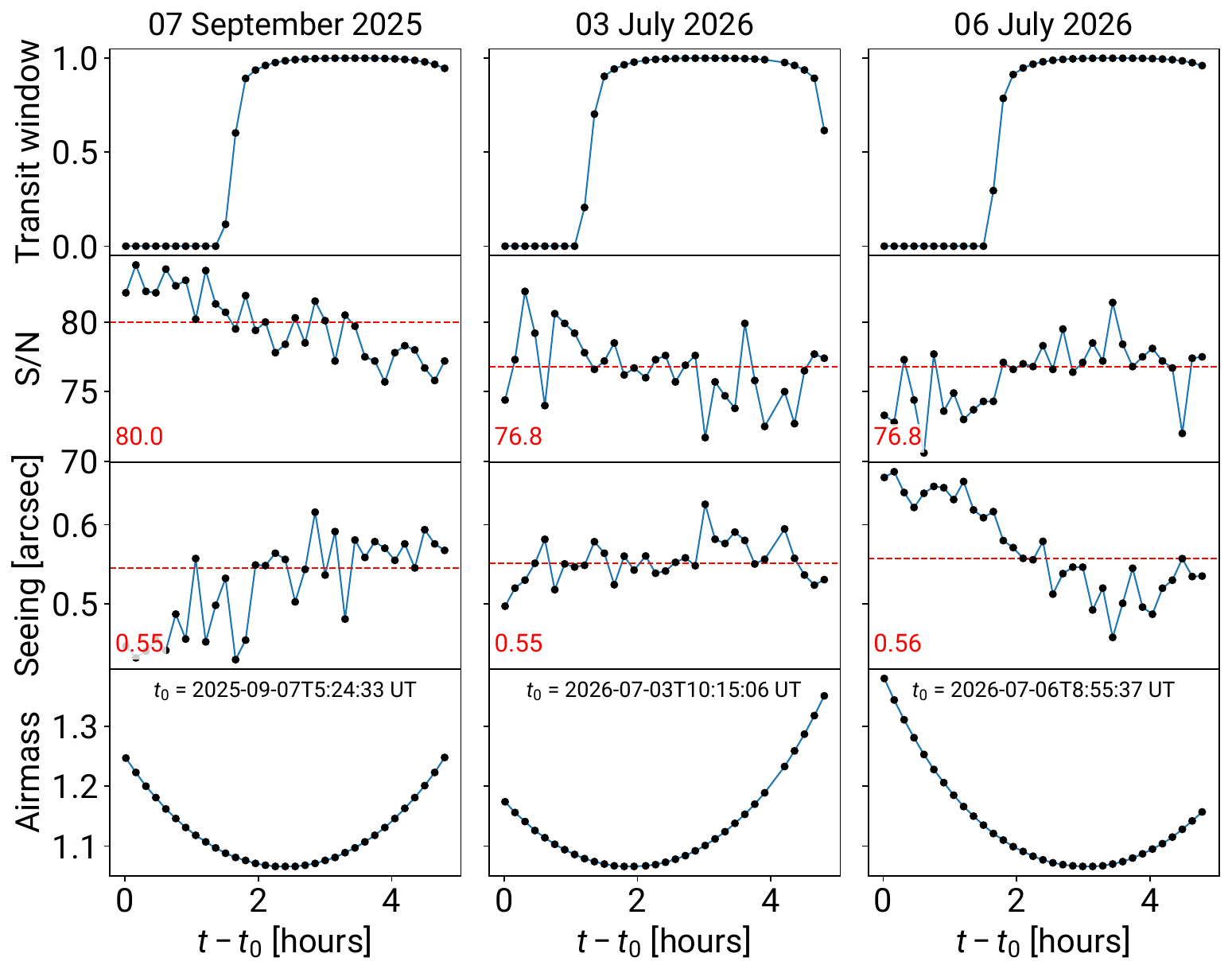}
   \caption{Evolution of the transit window, median S/N, seeing, and airmass for our three observed transits of KELT-9b, as a function of time since $t_0$, the start time of the first exposure of each night. One column is shown per transit. The red dashed lines indicate the median S/N and seeing over each observing sequence, with the median value given as a small red number in the lower left corner of each panel.}
   \label{fig:obslog}
\end{figure}

\begin{table}[!ht]
\caption{Log of the three SPIRou transit observations of KELT-9b.}
\label{tab:obslog}
\centering
\begin{tabular}{lccccccc}
\hline\hline
Date & Run ID & $N_{\rm seq}$ & $N_{\rm transit}$ & UT range & Airmass & Median S/N & Median seeing \\
 & & & & & & & [arcsec] \\
\hline
07 September 2025 & 25BP45 & 33 & 23 & 05:24--10:13 & 1.07--1.25 & 80.0 & 0.55 \\
03 July 2026      & 26AP45 & 32 & 24 & 10:15--15:04 & 1.07--1.35 & 76.8 & 0.55 \\
06 July 2026      & 26AP45 & 33 & 22 & 08:55--13:43 & 1.07--1.38 & 76.8 & 0.56 \\
\hline
\end{tabular}
\tablefoot{$N_{\rm seq}$ is the total number of polarimetric sequences acquired and $N_{\rm transit}$ the number of those whose mid-exposure BJD$_{\rm TDB}$ falls between first and fourth contact, from the ephemeris and geometry of Table~\ref{tab:system_parameters}. The UT range runs from the start of the first sub-exposure to the end of the last one. The S/N is the one estimated by the pipeline for each polarimetric sequence.}
\end{table}

\FloatBarrier
\section{$K_{\rm p}$--$v_{\rm sys}$ maps of the three transits}
\label{app:kpmapall}

\begin{figure}[!ht]
\sidecaption
\includegraphics[width=0.585\textwidth]{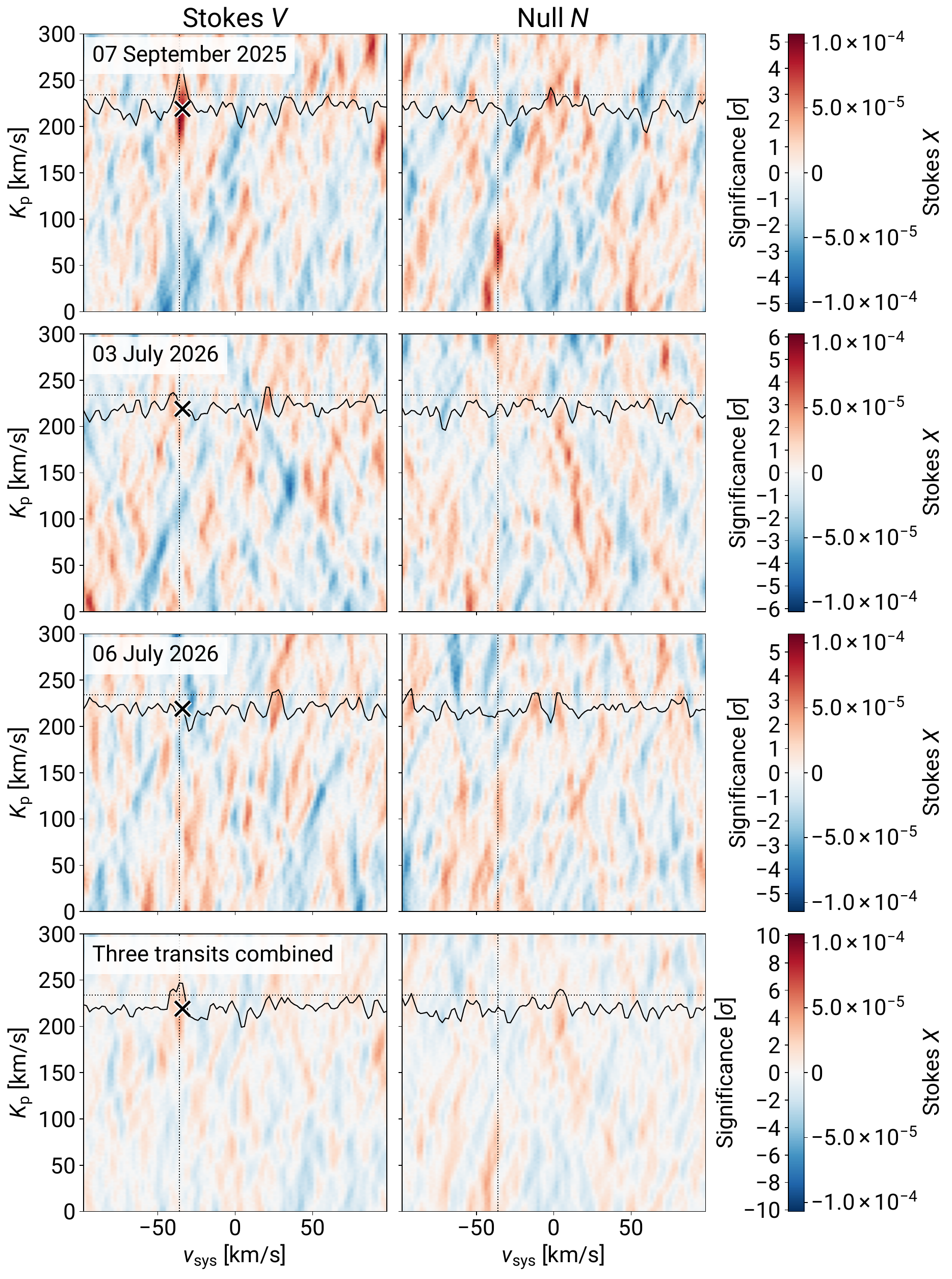}
 \caption{$K_{\rm p}$--$v_{\rm sys}$ maps of the transit-averaged Stokes~$V$ (left) and null~$N$ (right) LSD profiles, from top to bottom: 07 September 2025 (the map of Fig.~\ref{fig:kpmap}), 03 July 2026, 06 July 2026, and the three transits combined. The black curves are the profiles at $K_{\rm p} = 219$~km\,s$^{-1}$, drawn on one scale in every panel as in Fig.~\ref{fig:kpmap}. All four share a single colour scale in $V/I_c$, matched to Fig.~\ref{fig:kpmap}; each significance axis is in units of that map's own noise, which is lower for the combined map. The cross marks the 07 September 2025 signal ($K_{\rm p} = 219$~km\,s$^{-1}$, $v = -34$~km\,s$^{-1}$), repeated in the other panels. Neither 2026 night shows a signal at that position, and their strongest excursions elsewhere ($5.1\sigma$ and $4.7\sigma$ in absolute value) far exceed what Gaussian noise would give for a map this size (Sect.~\ref{sec:disc_variability}); the combined map reaches $3.2\sigma$ at that position.}
   \label{fig:kpmapall}
\end{figure}

\begin{figure}[!ht]
   \centering
   \includegraphics[width=0.558\textwidth]{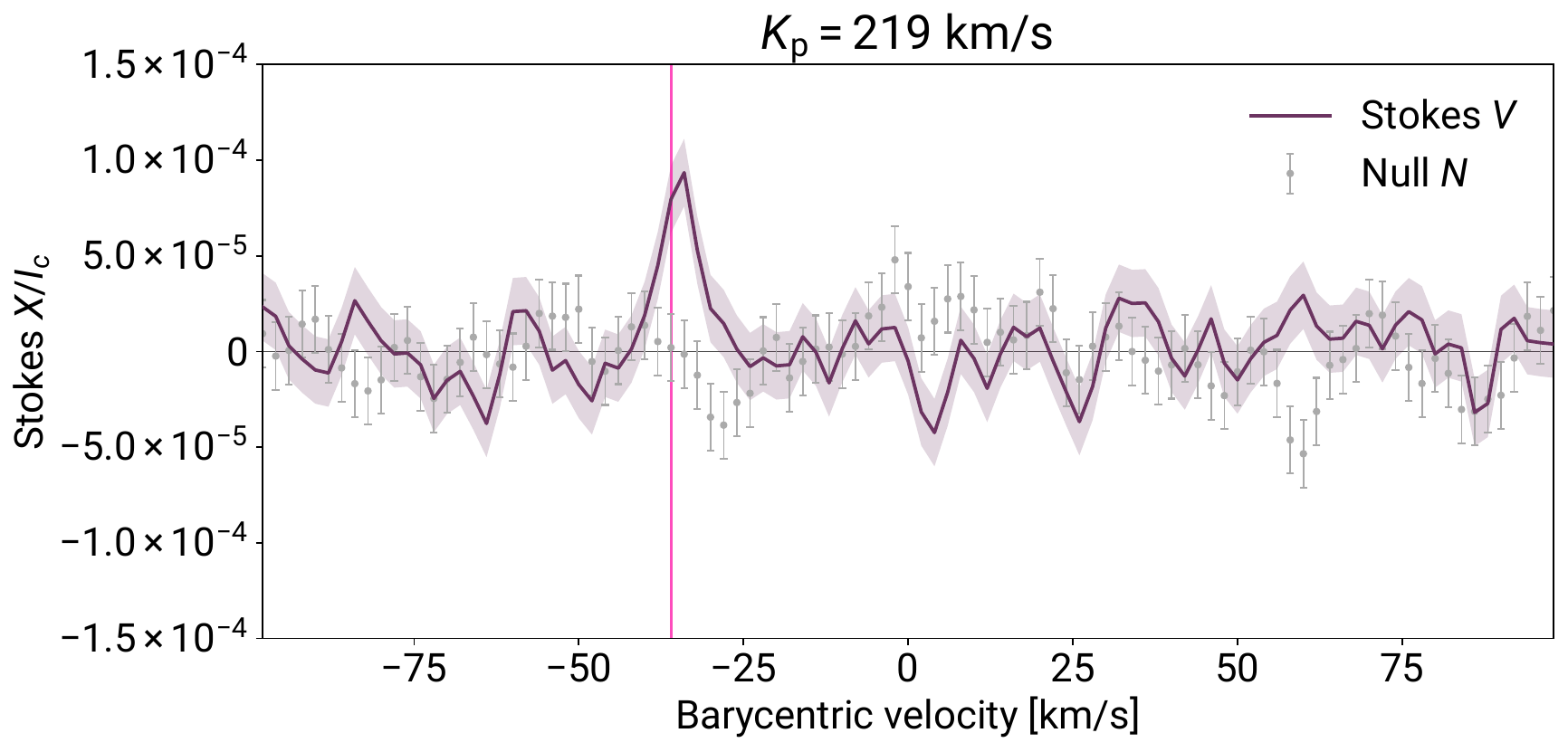}
   \caption{Stokes $V$ (solid purple line) and null $N$ (gray error bars) LSD profiles of KELT-9b at $K_{\rm p} = 219$~km\,s$^{-1}$, in the barycentric frame, for the transit of 07 September 2025. The vertical line marks $-36$~km\,s$^{-1}$, the velocity expected if a single limb contributes, from the systemic velocity, planetary rotation, and day-to-night wind (Sect.~\ref{sec:disc_velocity}), as in Fig.~\ref{fig:kpmap}.}
   \label{fig:Vprofile}
\end{figure}

Figure~\ref{fig:kpmapall} shows the maps of the three transits together, built identically and on a single colour scale, along with the map built from all 69 in-transit sequences at once. Fig.~\ref{fig:Vprofile} shows the LSD profiles extracted at the $K_{\rm p}$ of maximum detection on 07 September 2025. Combining the three transits leaves $3.2\sigma$ at the position of the 2025 signature, close to the $5.3/\sqrt{3}$ expected for a signature confined to a single epoch, and the strongest excursion of the combined map lies elsewhere ($4.4\sigma$, against $4.0\sigma$ in its null).

\section{Robustness of the 2025 detection}
\label{app:robustness}

The signature reported in Sect.~\ref{sec:results} was tested against the choice of data reduction and LSD code, and by three further tests. The observations were reduced twice and independently: with \texttt{APERO} (version 0.7.295) described in \citet{cook2022}, the standard SPIRou pipeline and the one adopted in the main text, and with \texttt{Libre-ESpRIT} \citep{donati1997}, an independent optimal-extraction and demodulation chain. The LSD step was checked in the same way: repeating the analysis with \texttt{specpolFlow} \citep{folsom2025} in place of our implementation of the algorithm of \citet{kochukhov2010} gives identical profiles and error bars, and maps agreeing to every digit of Table~\ref{tab:robustness}, so the deconvolution code contributes nothing to the difference between the two reductions.

\begin{figure}[!ht]
    \sidecaption
    \includegraphics[width=0.630\textwidth]{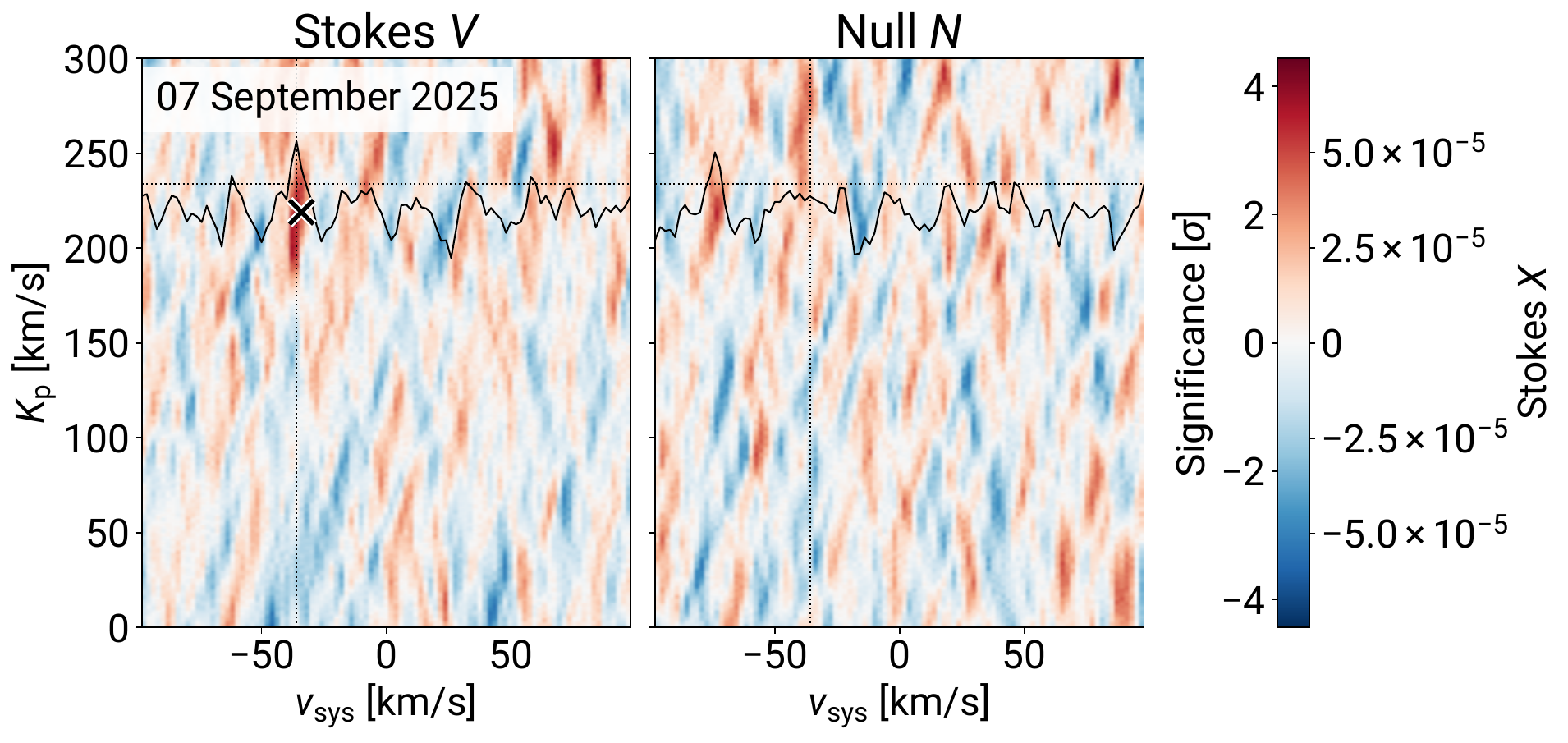}
   \caption{$K_{\rm p}$--$v_{\rm sys}$ maps of the transit-averaged Stokes~$V$ (left) and null~$N$ (right) LSD profiles for the 07 September 2025 transit, from the \texttt{Libre-ESpRIT} reduction. This is the same transit as Fig.~\ref{fig:kpmap}, which shows the \texttt{APERO} reduction, and the two maps are built identically. The signature appears at $K_{\rm p} = 221$~km\,s$^{-1}$, $v = -36$~km\,s$^{-1}$ and $4.4\sigma$.}
   \label{fig:kpmaplibre}
\end{figure}

Table~\ref{tab:robustness} lists the peak recovered from each reduction and Fig.~\ref{fig:kpmaplibre} shows the \texttt{Libre-ESpRIT} map. The position is stable ($K_{\rm p} = 219$ and $221$~km\,s$^{-1}$, $v = -34$ and $-36$~km\,s$^{-1}$); the significance differs more, $5.3\sigma$ against $4.4\sigma$, with a somewhat stronger null.

First, a signature produced in transit should be present only during transit. Repeating the analysis with all 33 sequences of the night instead of the 23 in transit lowers the Stokes~$V$ signature from $5.3\sigma$ to $5.0\sigma$ and its amplitude from $9.3$ to $7.6 \times 10^{-5}$. The 23 in-transit sequences span 3.3~h of the $T_{14} = 3.9$~h transit, which is therefore not fully covered.  Second, an independent split of the mask at its median wavelength (1704~nm) recovers the signature at $4.8\sigma$ from the red half, $V/I_c = (12.7 \pm 2.7) \times 10^{-5}$, and $1.9\sigma$ from the blue, $(4.3 \pm 2.3) \times 10^{-5}$. Third, the signature should be carried by the lines the planet is known to show. Restricting the mask to the 959 lines of \ion{Fe}{i}, a species robustly detected in KELT-9b \citep{hoeijmakers2018,borsato2026}, recovers $V/I_c = 9.9 \times 10^{-5}$ at the same position at $4.0\sigma$, against the $4.3\sigma$ expected from the retained mask weight.

We also investigated whether the signal could be of stellar origin, which we found to be unlikely.  A mask built for a $T_{\rm eff} = 10\,000$~K atmosphere with the criteria used for the planet retains 3 lines against the 1398 of the planetary mask. The 138-line stellar mask of Appendix~\ref{app:stellar} is larger by keeping lines deeper than 2\% of the continuum (instead of 10\%). A stellar signature would also fall at $K_{\rm p} = 0$. Additionally, the stellar LSD profiles show no Stokes~$V$ signature in the stellar rest frame (Appendix~\ref{app:stellar}).

The sensitivity of the analysis was checked by injecting synthetic signatures into the data before reduction and recovering them. In Stokes~$V$, we injected into each of the three transits a Gaussian of width $\sigma_v = 3$~km\,s$^{-1}$ at $K_{\rm p} = -230$~km\,s$^{-1}$, far from the signature, extending the map to $K_{\rm p} = -300$ to $300$~km\,s$^{-1}$ for this test, and varied its amplitude. The analysis returns $74\%$ of the injected peak in 2025, and $63\%$ and $64\%$ in 2026. A signal with an amplitude of $5 \times 10^{-5}$ is not detected, while $10^{-4}$ is recovered at $2.5$--$3.3\sigma$, and $5 \times 10^{-4}$ is detected on all three nights, $2 \times 10^{-4}$ on two.  At $9.3 \times 10^{-5}$ the 2025 signature therefore sits close to the detection threshold of the method. The same test in Stokes~$I$, on the 2025 transit, sets an upper limit of $3 \times 10^{-4}$ on any planetary signature. Our detection limit is weaker for Stokes $I$ as we need to run an additional step in the analysis: the removal of the stellar signal using principal component analysis.

The significance in units of $\sigma$ quoted in Sect.~\ref{sec:results} is computed from a single pixel of the map. We therefore also computed the FAP of the signature  from the $\chi^2$ of the LSD profile over $|v+34| \leq 9$~km\,s$^{-1}$, the $\pm3\sigma$ width of the fitted Gaussian. The 2025 Stokes~$V$ signature reaches $\mathrm{FAP} = 1.9\times10^{-4}$. The two 2026 transits are non-detections (FAP $0.54$ and $0.19$). 

\begin{table}[!ht]
\caption{Peak recovered for the 07 September 2025 transit from each data reduction.}
\label{tab:robustness}
\centering
\begin{tabular}{lcccc}
\hline\hline
Reduction & Peak & $K_{\rm p}$ & $v$ & max\,$|N|$ \\
 & [$\sigma$] & [km\,s$^{-1}$] & [km\,s$^{-1}$] & [$\sigma$] \\
\hline
\texttt{APERO}        & 5.3 & 219 & $-34$ & 3.9 \\
\texttt{Libre-ESpRIT} & 4.4 & 221 & $-36$ & 4.1 \\
\hline
\end{tabular}
\tablefoot{max\,$|N|$ is the strongest excursion of the null map of the same analysis, scaled by the Stokes~$V$ noise.}
\end{table}

\FloatBarrier
\section{The stellar magnetic field of KELT-9}
\label{app:stellar}

\begin{figure}[!ht]
   \centering
   \includegraphics[width=0.765\linewidth]{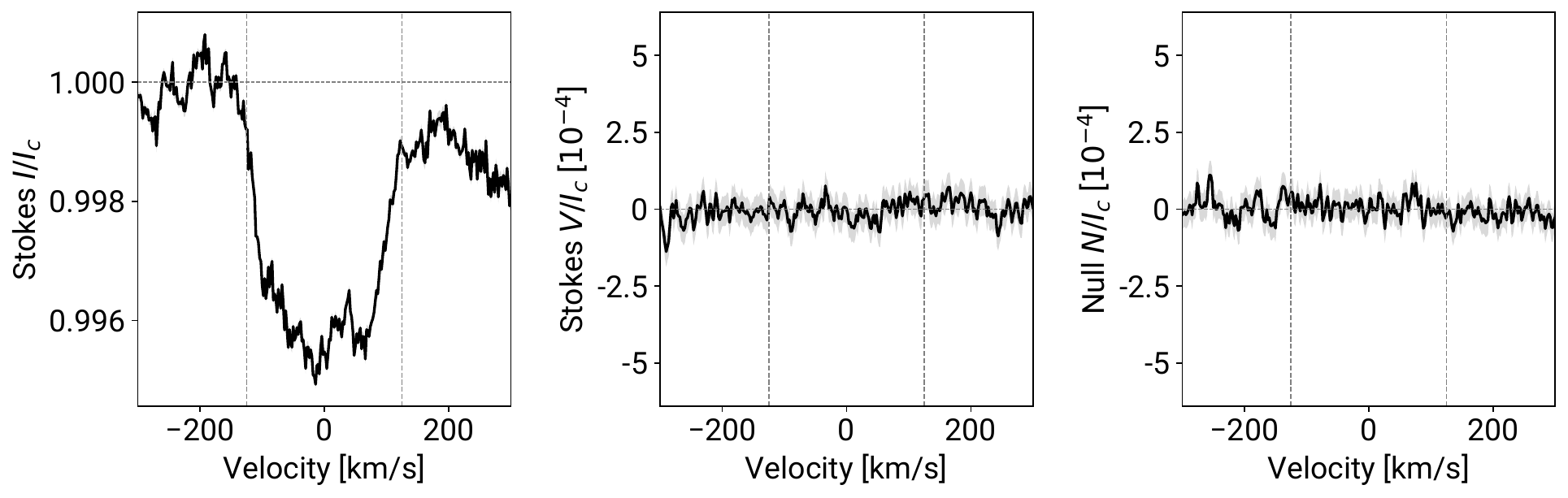}
   \caption{Stellar mean LSD profiles of KELT-9 in the stellar rest frame: Stokes~$I$ (left), Stokes~$V$ (middle), and null~$N$ (right), with $1\sigma$ error bands. The dashed lines mark the integration range for the mean longitudinal magnetic field.}
   \label{fig:stellarlsd}
\end{figure}

The stellar spectra can be used to search for a magnetic field at the surface of the star. We therefore computed stellar LSD profiles using a mean stellar spectrum computed with all 98 sequences of the three nights (both in transit and out of transit). The line mask was extracted from VALD3 with $T_{\rm eff} = 10\,000$~K, close to that of the star, and $\log g = 3.5$. We kept the atomic lines with a known effective Land\'e factor and deeper than 2\% of the continuum, excluding H, He, Na, K, and Ca, as for the planetary mask. It contains 138 lines with a mean depth of 0.06, a mean effective Land\'e factor of 1.17, and a mean wavelength of 1304~nm. The resulting profiles are shown in Fig.~\ref{fig:stellarlsd}. The Stokes~$I$ profile shows the rotationally broadened line expected for the star with its high projected rotational velocity. The Stokes~$V$ and null profiles are flat. 

We measured the mean longitudinal field from the first moment of Stokes~$V$  integrated over $|v| \leq 125$~km\,s$^{-1}$. We report a non-detection of the stellar magnetic field, with a $3\sigma$ upper limit of $|B_\ell| < 350$~G. The stellar lines show no circular polarization in the stellar rest frame, and a stellar signature would in any case not move with the planet.

\section{Forward toy-model of the transmission signature}
\label{app:forward}

\begin{figure}[!ht]
   \centering
   \raisebox{0.0687\linewidth}{\includegraphics[width=0.238\linewidth]{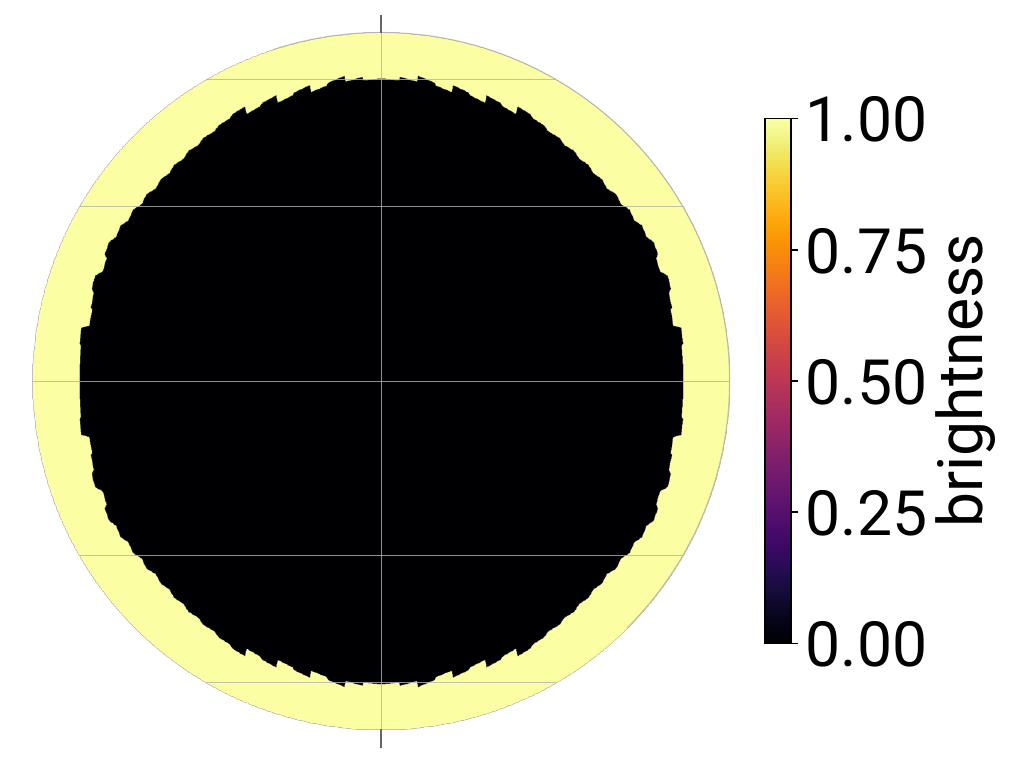}}%
   \raisebox{0.0687\linewidth}{\includegraphics[width=0.238\linewidth]{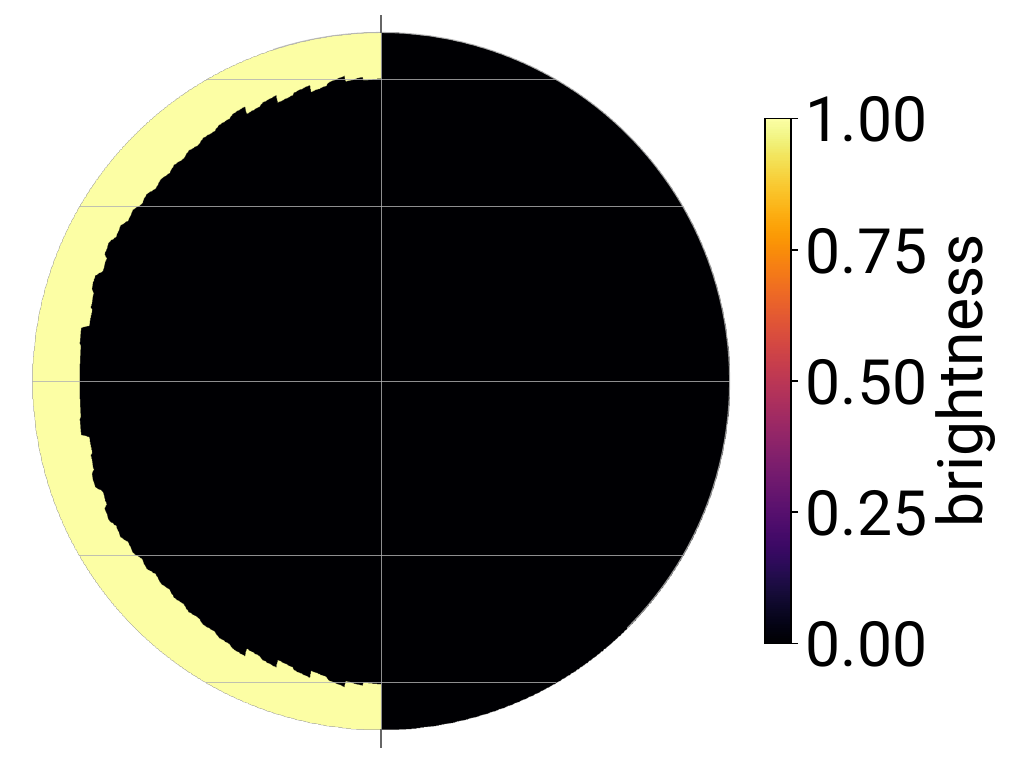}}%
   \includegraphics[width=0.379\linewidth]{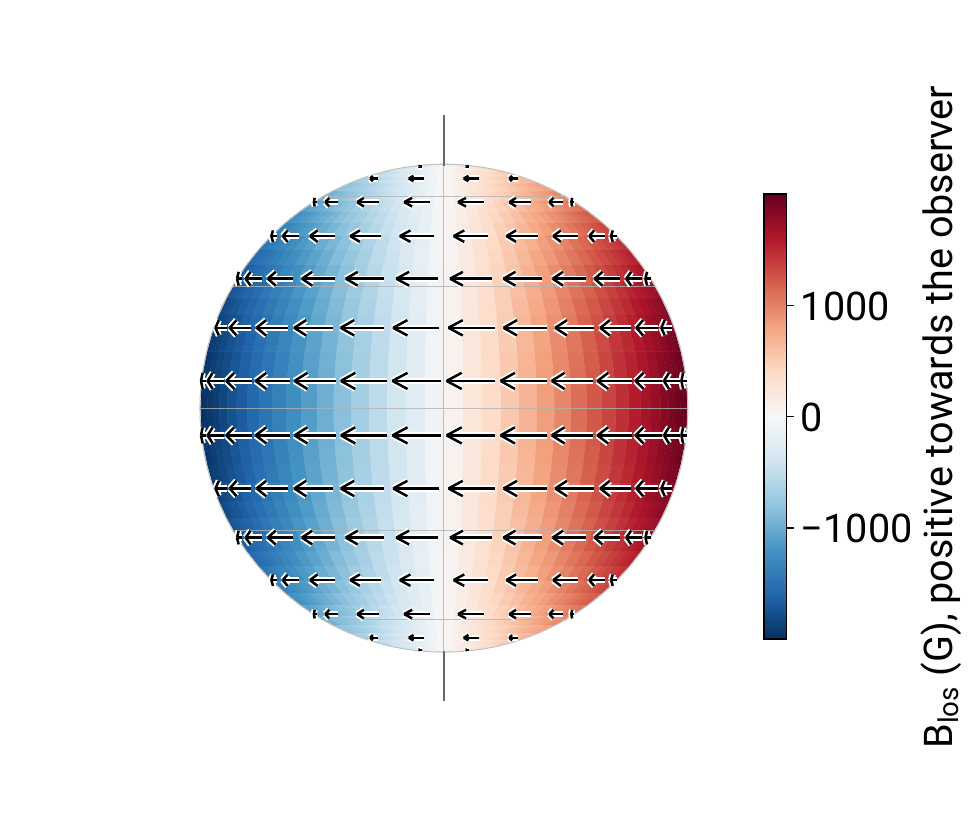}
   \caption{Input maps of the forward model, seen by the observer. Left and middle: the brightness maps with both limbs lit and with a single limb lit. Right: the toroidal magnetic field, coloured by its line-of-sight component and with arrows showing the field vectors, for $B_{\rm eq} = 2$~kG at the equator (a surface mean of 1.6~kG); the field is identical in the two configurations, which differ only in the brightness map.}
   \label{fig:fwdmaps}
\end{figure}

The forward toy-models of Sect.~\ref{sec:disc_scenarios} were computed with \texttt{ZDIpy} \citep{folsom2018zdipy}, which integrates local Voigt profiles over a rotating sphere. We replaced its weak-field treatment by the Unno--Rachkovsky solution for a Zeeman triplet, since the Zeeman splitting below is not small against the line width. The planet is assumed to be a sphere seen equator-on ($i = 90\degr$) rotating at $v_{\rm rot,p} = 6.4$~km\,s$^{-1}$ \citep{zhang2026}, divided into 4584 cells. The transmission annulus is imposed as a brightness map: a ring of constant angular thickness around the projected disc contributes, lit uniformly (Fig.~\ref{fig:fwdmaps}, left). The local line has the mean wavelength and Land\'e factor of the planetary mask ($1747$~nm, $g_{\rm eff} = 1.27$), a Gaussian width of 2.4~km\,s$^{-1}$, and a depth scaled so that the disc-integrated Stokes~$I$ profile is $10^{-4}$ deep; profiles are convolved to $R = 70\,000$.

\begin{figure}[!ht]
   \sidecaption
   \includegraphics[width=0.630\textwidth]{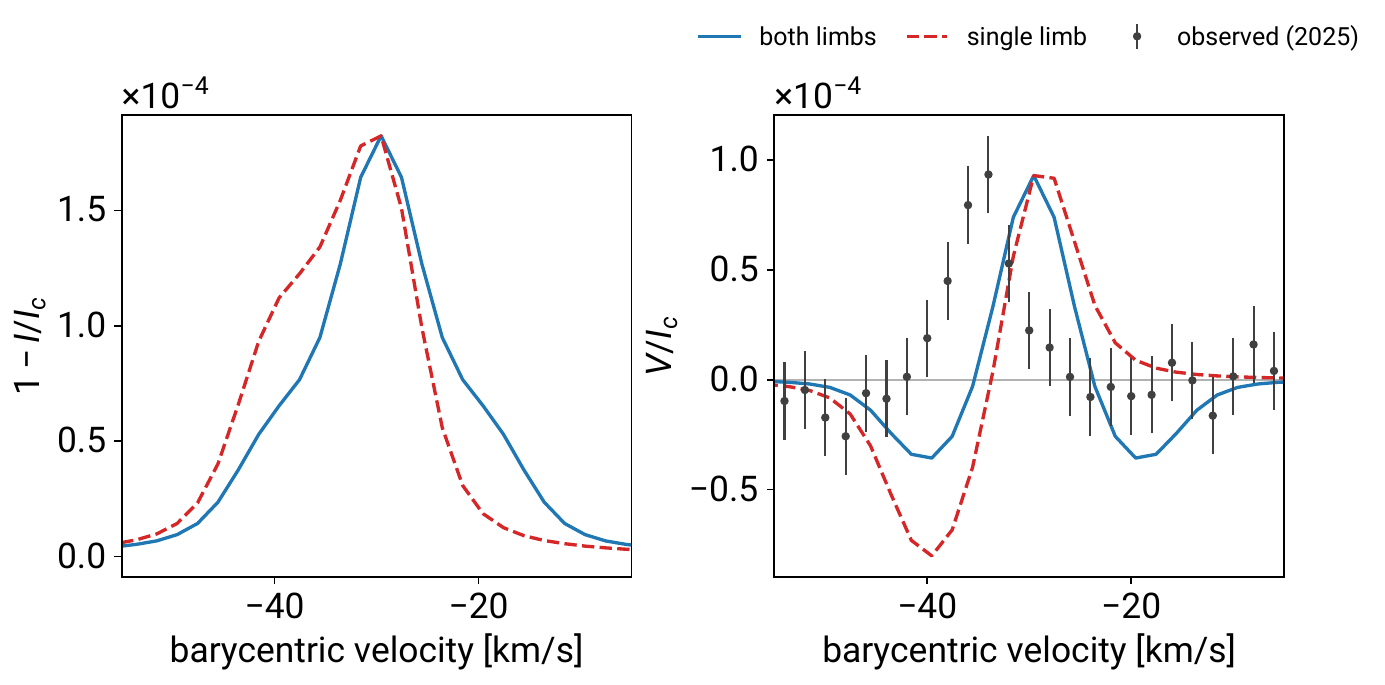}
   \caption{Forward toy-models with both limbs lit (solid blue) and a single limb lit (dashed red), for the same magnetic map with Stokes~$I$ profiles on the left panel, and Stokes~$V$ profiles on the right panel compared with the observed signature. }
   \label{fig:fwdlimbs}
\end{figure}

The field is purely toroidal and axisymmetric ($\ell = 1$, $m = 0$), directed towards the observer on one limb and away on the other (Fig.~\ref{fig:fwdmaps}, right). Fig.~\ref{fig:fwdlimbs} shows the resulting profiles for $B_{\rm eq} = 2$~kG, a surface mean of 1.6~kG. At the equator this field splits the mean line by 6.2~km\,s$^{-1}$, 2.6 times its Gaussian width, which is where Stokes~$V$ is strongest: it reaches $6.2 \times 10^{-5}$ of the continuum there and weakens for stronger fields, as the two lobes separate at constant area. To match the $V/I_c = 9.3 \times 10^{-5}$ measured in 2025, we need to scale the line depth so that the intensity profile is $1.8 \times 10^{-4}$ deep rather than $10^{-4}$.

We repeated the calculation with the same field and weighting but a single limb lit, to test the asymmetric geometry of Sect.~\ref{sec:disc_velocity}. Fig.~\ref{fig:fwdlimbs} shows the two models with the observed profile. Each model is scaled so that its strongest lobe reaches $V/I_c = 9.3 \times 10^{-5}$, and the left panel shows the Stokes~$I$ profile each geometry then requires. Both model spectra are formally inconsistent with the data.

\end{appendix}

\end{document}